\documentclass[aps,prb,amsmath,amssymb,floatfix,twocolumn,eqsecnum]{revtex4-2}
\usepackage{graphicx}
\usepackage{dcolumn}
\usepackage{bm}
\usepackage{hyperref}
\usepackage{braket}
\usepackage{shorthand}
\usepackage{makecell}
\usepackage{xcolor}

\begin{document}
\title{Intertwined Order in Fractional Chern Insulators from Finite-Momentum Pairing of Composite Fermions}
\author{Ramanjit Sohal}
\author{Eduardo Fradkin}
\affiliation{Department of Physics and Institute for Condensed Matter Theory,
University of Illinois at Urbana-Champaign, 1110 West Green Street, Urbana, Illinois 61801, USA}
\begin{abstract}
We investigate the problem of intertwined orders in fractional Chern insulators by considering lattice fractional quantum Hall (FQH) states arising from pairing of composite fermions in the square-lattice Hofstadter model. At certain filling fractions, magnetic translation symmetry ensures the composite fermions form Fermi surfaces with multiple pockets, leading to the formation of finite-momentum Cooper pairs in the presence of attractive interactions. We obtain mean-field phase diagrams exhibiting a rich array of striped and topological phases, establishing paired lattice FQH states as an ideal platform to investigate the intertwining of topological and conventional broken symmetry order. 
\end{abstract}
\date{\today}

\maketitle

\section{Introduction}

Fractional quantum Hall (FQH) states realized in lattice systems have attracted considerable attention in recent years, driven in large part by advances in the engineering of Chern bands in solid-state Moir\'e \cite{Spanton2018,Cao2018a,Cao2018b,Sharpe2019,Chen2020,Serlin2019} and cold atom systems \cite{Aidelsburger2013,Miyake2013,Aidelsburger2014,Jotzu2014,Cooper2019}. In the presence of strong interactions, the partial filling of a Chern band may result in the formation of a fractional Chern insulator (FCI) state \cite{Neupert2011,Sheng2011,Tang2011,Sun2011,Regnault2011,Bergholtz2013,Parameswaran2013}, a lattice analogue of continuum FQH states \cite{Laughlin1983}. 
Importantly, lattice effects can give rise to phenomena with no continuum analogue, such as novel FCI states  obtained by partial filling of bands with Chern number greater than unity, which may support lattice defects with non-trivial braiding statistics \cite{Sterdyniak2013,Barkeshli2012}.
The presence of the lattice also results in the competition and, in some cases, the coexistence of FCI states with more traditional broken symmetry orders, such as charge density waves (CDWs) \cite{Kourtis2014,Kourtis2017}. 
This phenomenon of multiple orders that sometimes compete with each other but sometimes drive each other is reminiscent of the complex intertwined orders found in high temperature superconductors \cite{Berg2009,Fradkin2015}.

In spite of the importance of the lattice, many FCI states can still be understood through the widely used composite fermion (CF) framework \cite{Jain1989,Lopez1991}, like most experimentally observed continuum FQH states. In this picture, the electrons nucleate fluxes of an emergent Chern-Simons gauge field, which partially screen the external magnetic field. The bound states of the electrons and the emergent flux are known as composite fermions. In the continuum, a FQH state of electrons results when the composite fermions, which feel a reduced net flux, form an integer quantum Hall state. 
Much as in the case of the continuum FQH states, the FCI lattice counterparts can also be represented in terms of a theory of (composite) lattice fermions coupled to a lattice version of Chern-Simons gauge theory \cite{Fradkin1989,Eliezer1992,Eliezer-1992b,Sun2015}.
Abelian FCI states are formed when composite fermions fill an integer number of Hofstadter bands \cite{Kol1993,Moller2009,Moller2015,Sohal2018}. At certain filling fractions in the continuum case, the composite fermions see no effective flux and so form a Fermi surface \cite{Halperin1993}. In higher Landau levels, this composite Fermi liquid yields to a pairing instability, resulting in a $p_x+ip_y$ superconductor of composite fermions \cite{Read2000}. This gapped state is the Pfaffian state proposed by Moore and Read \cite{Moore1991} and posseses non-Abelian topological order. 

Although analogues of the Pfaffian state have been observed numerically in lattice systems \cite{Wu2012,Wang2012,Liu2013a,Liu2013b,Wang2015}, we claim that more exotic paired phases may also be obtainable. Indeed, although the composite fermions may form a Fermi surface at certain filling fractions due to the vanishing of the net flux, at other filling fractions at which the net flux is non-zero, the composite fermions may partially fill a Hofstadter band and so still form a Fermi surface.
Magnetic translation symmetry implies, as we will review, that this Fermi surface must consist of multiple Fermi pockets, raising the possibility of finite-momentum pairing and the formation of Fulde-Ferrell-Larkin-Ovchinnikov (FFLO) \cite{Fulde1964,Larkin1964} or pair-density wave (PDW) \cite{Berg2009} like states. These statements may hold true even in zero magnetic field, as the composite fermions will still see a non-zero Chern-Simons flux. We should emphasize that the PDW states we investigate do not arise from a Zeeman effect (as in the conventional FFLO states) but rather have a purely orbital origin.

The goal of the present study is to illustrate the existence, at mean-field level, of a novel set of FCI phases which exhibit a coexistence of topological order (TO) and broken symmetry order (BSO) as a result of finite-momentum composite fermion pairing, taking as an example, for simplicity, the square-lattice Hofstadter model \footnote{See Refs. \cite{Zhai2010,Iskin2015,Umucalilar2016,Guo2018} for related studies of pairing of electrons in the spin-$1/2$ Hofstadter model}. We find, for instance, topologically ordered states supporting CDWs, providing a new entry in the long history of stripe order in QH systems \cite{Fogler1996,Moessner1996,Koulakov1996,Fradkin1999,Fradkin2000}. These states support a range of Abelian and non-Abelian topological orders, including the Pfaffian and PH-Pfaffian \cite{Son2015} states. We also find a phase which we call a quantum Hall thermal semimetal, as the charge sector is gapped, while the neutral sector is described by a theory of relativistic massless Majorana fermions.
Such a state will possess a quantized Hall conductance, but will support unquantized transport of heat through the bulk.

Related phenomena have been exhibited in recent experiments \cite{Samkharadze2016,Schreiber2017,Schreiber2018}, which revealed a competition between pairing and nematicity in continuum Landau levels. A subsequent theoretical study \cite{Santos2019} proposed a $p_x+ip_y$ PDW state of composite fermions as a possible explanation for this observation. The distinguishing feature between the physics we present and that of, for instance, Ref. \cite{Santos2019} is that we present FCIs as a platform in which to study \emph{intertwining} of TO and BSO, in that they do not compete with nor are even independent of one another, but rather arise from a common microscopic origin, namely the interplay between the pairing of composite fermions and the commensurability of the lattice and magnetic length scales. 

We emphasize that, although we focus on a particular lattice model, the basic mechanism of finite-momentum pairing of composite fermions is applicable to other experimentally relevant lattice systems. These include the aforementioned Moir\'e systems, such as bilayer graphene/hexagonal boron nitride heterostructures, in which Abelian fractionalized states have been observed in strong magnetic fields \cite{Spanton2018}. Recent theoretical studies suggest that such states may even be found at zero magnetic field in twisted bilayer graphene systems \cite{Abouelkomsan2020,Ledwith2020,Repellin2020,Liu2020}. On the cold atoms front, the Hofstadter model has already been experimentally realized \cite{Aidelsburger2013,Miyake2013,Aidelsburger2014}. Although fractionalized states have not yet been observed, the tunability of interactions in these systems make them a promising playground in which to search for our proposed finite momentum paired states. With this in mind, we look for both \emph{fermionic and bosonic }FCI states in the Hofstadter model, the latter of which are of relevance to cold atom experiments. At the filling fractions we consider, the bosonic and fermionic phase diagrams exhibit roughly the same set of ordered states.

The remainder of this paper is structured as follows. First, we introduce the fermionic Hofstadter model and review the flux attachment transformation. We identify three example filling fractions at which the composite fermions form Fermi surfaces with multiple Fermi pockets. Next, we perform a self-consistent BCS calculation to produce phase diagrams at these fillings in the presence of attractive nearest-neighbor (NN) and repulsive next-nearest-neighbor (NNN) interactions. We then briefly repeat this analysis for the same lattice model, but with hardcore bosons. Lastly, we discuss our results and conclude.

\section{Model, Flux Attachment, and Compressible FCI States}

We consider the Hofstadter model \cite{Harper1955,Azbel1964,Hofstadter1976} of spinless fermions hopping on a square lattice in a uniform magnetic field, as described by the Hamiltonian
\begin{align}
	H_0 = -t\sum_{\bm x} \sum_{j=x,y} \left[ c_{\bm x}^\dg c_{\bm x + \bm e_j}^{\pdg} e^{-iA_j(\bm x)} + H.c. \right], \label{eqn:hofstadter-hamiltonian}
\end{align}
where $t$ is the hopping amplitude, $\bm e_j$ are the NN lattice vectors, and $A_j(\bm x)$ is the electromagnetic vector potential. We choose the Landau gauge $\bm A = (0, \phi_0 x)$, where $\phi_0$ is the flux per plaquette. We take
\begin{align}
\phi_0 = 2\pi \frac{p_0}{q_0}, \quad p_0,q_0 \in \mathbb{Z},
\end{align}
with $p_0$ and $q_0$ co-prime, so that the magnetic unit cell (MUC) consists of $q_0$ sites along the $x$ direction. The energy spectrum therefore consists of $q_0$ bands. Additionally, the magnetic translation algebra \cite{Zak1964} dictates that the single particle dispersion obeys the following periodicity in the magnetic Brillouin zone (MBZ):
\begin{align}
	\varepsilon(k_x , k_y) = \varepsilon(k_x, k_y - \phi_0). \label{eqn:dispersion-mts-periodicity}
\end{align}
The consequences of magnetic translation symmetry will play an important role when we turn to discussing pairing of composite fermions. 
 
Now, the Chern number, $C_0$, of the first $r$ filled bands of the Hofstadter Hamiltonian satisfies the Diophantine equation $r = C_0 p_0 + D_0 q_0, \quad D_0 \in \mathbb{Z}$ \cite{Thouless1982}. The lowest Landau level (LLL) corresponds to the solution $(r,C_0,D_0)=(p_0,1,0)$. Hence, lattice effects split the LLL into $p_0$ sub-bands. We are interested in scenarios in which the LLL filling $\nu \equiv 2\pi n / \phi$, where $n$ is the fermion density per site, is fractional. Here we are following the conventions of Ref. \cite{Moller2015} by defining the filling relative to the bands below a certain gap (in this case, the gap above the manifold of states corresponding to the LLL), rather than in terms of the filling of a specific band.

\begin{figure}
  \includegraphics[width=0.27\textwidth]{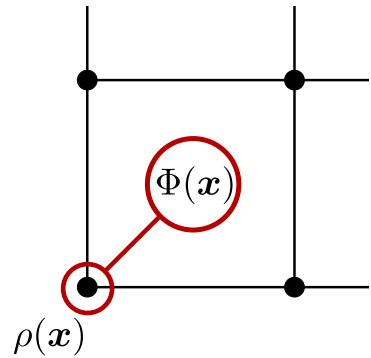}
  \caption{Flux attachment on the square lattice. The Chern-Simons flux, $\Phi(\bm x)$, through the plaquette north-east of $\bm x$ is attached to the fermion density $\rho(\bm x)$ via Gauss' law, $\rho(\bm x) = \theta \Phi(\bm x)$.}
  \label{fig:flux-attachment}
\end{figure}

We look for fractionalized phases at these filling fractions by performing an exact mapping of the system of fermions to a system of composite fermions coupled to an emergent Chern-Simons gauge field \cite{Fradkin1989,Lopez1991}. Physically speaking, this flux attachment procedure amounts to attaching solenoids of $2k$, $k\in \mathbb{Z}$, flux quanta to each fermion so that the resulting bound state of a fermion and a solenoid, a composite fermion, still obeys Fermi statistics.
The resulting action is given by
\begin{align}
S[f,f^\dg,a_\mu] = S_F[f,f^\dg,a_\mu] + S_{CS}[a_\mu] 
\end{align}
where $f$ is the composite fermion field and $a_\mu$ the statistical gauge field. 
Explicitly,
\begin{align}
\begin{split}
&S_F = \int_t \sum_{\bm x} \left[ f^\dagger (\bm x,t)(iD_0 + \mu)f(\bm x,t) \right. \\
+ & \sum_{j=x,y} \left. (f^\dg(\bm x,t)e^{i (a_j(\bm x,t) - A_{j}(\bm x))} f(\bm x+ \bm e_j,t) + H.c. ) \right],
\end{split}
\end{align}
where $D_0 = \partial_0 + ia_0$ is the covariant time derivative and $\mu$ is the chemical potential. 
The flux attachment procedure on the lattice is more subtle than that in the continuum due to the difficulties associated with defining a lattice Chern-Simons action. We make use of the action defined in Refs. \cite{Eliezer1992,Sun2015}, which takes the form,
\begin{align}
\begin{split}
S_{CS} = \theta \int_t &\sum_{\bm x} \left[ a_0 (\bm x,t) \Phi(\bm x,t)  - \frac{1}{2} a_i(\bm x,t) \mathcal{K}_{ij} \dot{a}_j(\bm x,t) \right] .  \end{split} \label{eqn:SMdiscCS}
\end{align}
Here,
\begin{align}
\theta = 1 / 2\pi (2k), \, k \in \mathbb{Z}
\end{align}
and $\Phi(\bm x) \equiv \epsilon_{ij} d_i a_j(\bm x)$ is the Chern-Simons flux through the plaquette north-east of the site $\bm x$, where the $d_i$ are forward difference operators: $d_i a_j(\bm x) = a_j(\bm x + \bm e_i) - a_j(\bm x)$. Likewise, we define backward difference operators, $\hat{d}_i$, which have the action, $\hat{d}_i a_j(\bm x) = a_j(\bm x) - a_j(\bm x - \bm e_i)$. The operator $\mathcal{K}_{ij}$ -- the explicit form of which is unimportant for us and is relegated to Appendix \ref{sec:appendix-fluc-attachment-details} -- is chosen so as to make the theory gauge-invariant. What is important is that $S_{CS}$ enforces the flux attachment constraint (or Gauss' law), $f^\dg f(\bm x) = \theta\Phi(\bm x)$, via the Lagrange multiplier field $a_0$, as depicted in Fig. \ref{fig:flux-attachment}.

We will defer the inclusion of interaction terms until the next section, as we first simply wish to understand the mean-field composite fermion band structure.
Now, the saddle-point equations for the above action are given by (restricting to time-invariant solutions)
\begin{align}
\la f^\dg(\bm x)f(\bm x) \ra \equiv \rho(\bm x) &= \theta \Phi(\bm x) \\
\la j_k(\bm x) \ra &=  \theta \epsilon_{ki} \hat{d}_i a_0(\bm x) 
\end{align}
where $j_k(\bm x) \equiv -\frac{\partial S_F}{\partial a_k(\bm x)}$ is the gauge-invariant current. On the square lattice, there always exists a uniform solution at any filling fraction with
\begin{align}
\rho(\bm x) \equiv n, \quad \Phi(\bm x) \equiv \phi  = n \theta, \quad j_k(\bm x) = a_0(\bm x) = 0.
\end{align}
In this mean-field configuration the composite fermions feel a reduced effective flux of
\begin{align}
\phi_* = \phi_0 - \phi \equiv 2\pi \frac{p_*}{q_*}
\end{align}
per plaquette, where we restrict ourselves to cases where $p_*$ and $q_*$ are integer and take them to be co-prime. So, the mean-field CF band structure is described by a Hofstadter Hamiltonian in the form of Eq. (\ref{eqn:hofstadter-hamiltonian}), but with a flux per plaquette of $\phi_*$. 

\begin{figure}
  \includegraphics[width=0.49\textwidth]{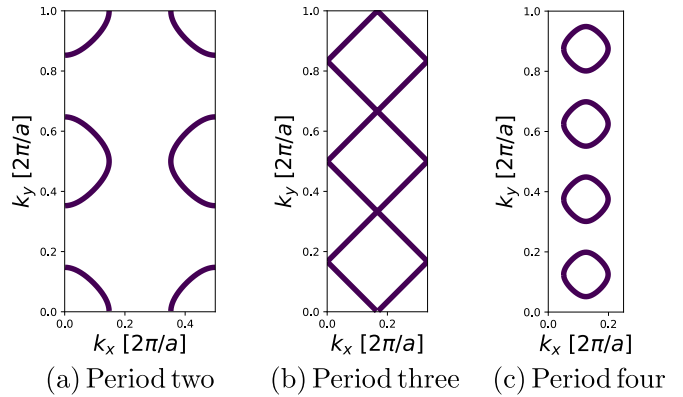}    
  \caption{Composite Fermi surfaces for the period two, three, and four configurations given in Table \ref{tab:fillings}.}
  \label{fig:FermiSurface}
\end{figure}

For appropriate choices of $\nu$ and $k$, the resulting mean-field spectrum consists of CFs partially filling a Hofstadter band, yielding a Fermi surface and hence a compressible state. In particular, if there is a CF pocket centered at, say, $\bm k = 0$, then magnetic translation symmetry implies, through Eq. (\ref{eqn:dispersion-mts-periodicity}), that there will be $q_*-1$ additional CF pockets centered at momenta $\bm Q_l = (0, \, 2\pi l / q_*)$, $l \in \mathbb{Z}$, in the Landau gauge. This is illustrated in Fig. \ref{fig:FermiSurface} for the three different configurations of magnetic flux and filling specified in Table \ref{tab:fillings}. Given the number of Fermi pockets for each configuration, we will label them as period two, three, and four, respectively. It is clear that, in the presence of an attractive interaction, we have the possibility of the formation of Cooper pairs of CFs with center of mass momenta $\bm Q_l + \bm Q_m$. 

\begin{table}
\caption{\label{tab:fillings}%
Details of the three composite Fermi liquid states whose pairing instabilities we investigate. The names period two, three, and four refer to the periodicity of the MBZ. Here $\phi_0$, $n$, $\nu$, $k$, $\phi$, and $\phi_*$ are the magnetic flux, fermion density per site, LLL filling fraction, number of pairs of attached statistical flux quanta, statistical flux, and effective flux seen by the composite fermions.
}
\begin{ruledtabular}
\begin{tabular}{ccccccc}
 &
$\phi_0/2\pi$&
$n$ &
$\nu$ &
$k$ &
$\phi/2\pi$ &
$\phi_*/2\pi$\\
\colrule
Period two & $3/4$ & $1/8$ & $1/6$ & $1$ & $1/4$ & $1/2$ \\
Period three & $2/3$ & $1/6$ & $1/4$ & $1$ & $1/3$ & $1/3$ \\
Period four & $5/8$ & $3/16$ & $3/10$ & $1$ & $3/8$ & $1/4$
\end{tabular}
\end{ruledtabular}
\end{table}

\section{Mean-field Theory of Paired States}

Our goal now is to investigate the possible pairing instabilities when the composite fermions form a Fermi surface with multiple Fermi pockets, focusing, for simplicity, on the three configurations listed in Table \ref{tab:fillings}. To that end, we introduce a NN attractive interaction,
\begin{align}
\notag S_{\mathrm{pair}} &= - V \int_t  \sum_{\bm x,j} f^\dagger (\bm x,t) f^{\dagger}(\bm x+\bm e_j,t) f(\bm x+\bm e_j,t)f(\bm x,t) \\
\notag 		&\sim - \int_t \sum_{\bm x,j} \left[ \Delta_{\bm x,j} f^\dg(\bm x,t) f^\dg(\bm x+\bm e_j,t) \right.  \\
		&\qquad \left. + \Delta_{\bm x,j}^\dg f(\bm x+\bm e_j,t) f(\bm x,t) - \frac{1}{V}|\Delta_{\bm x,j}|^2 \right]
\end{align}
where $V<0$ and we have performed a Hubbard-Stratonovich transformation to introduce the complex pair field $\Delta_{x,j}$. We will also consider the effect of NNN repulsive interactions,
\begin{align}
\notag S_{\mathrm{int}} &= -\frac{g}{2}\int_t \sum_{\bm x, \bm x'} f^\dg(\bm x, t) f(\bm x , t) U(\bm x - \bm x') f^\dg (\bm x', t) f (\bm x ', t) \\
\notag			&\sim - g \int_t  \sum_{\bm x, \bm x'} \left[ f^\dg(\bm x, t) f(\bm x , t) U(\bm x - \bm x') \rho(\bm x ' , t) \right. \\
			& \qquad\qquad\qquad \left. - \frac{1}{2} \rho(\bm x , t) U(\bm x - \bm x') \rho(\bm x' , t) \right]
\end{align}
where $g>0$, $\rho(\bm x)$ is a Hubbard-Stratonovich field corresponding to the fermion density, and  $U(\bm x - \bm x')= 1 $ if $\bm x$ and $\bm x'$ are next-nearest-neighbors while $U(\bm x - \bm x')=0$ otherwise. We include this repulsive interaction in order to stabilize additional striped solutions, which may be metastable at $g=0$. Such a combination of short-range attractive and long-range repulsive interactions can be be engineered in cold atom systems and has been shown numerically to be conducive to the formation of non-Abelian FCI states \cite{Wang2015}. We will restrict our attention to the region of phase space in which $0 \leq g < -V$. 

Now, in principle, we could perform a fully self-consistent calculation and solve the saddle-point equations for the Hubbard-Stratonovich fields and the Chern-Simons gauge fields. Indeed, the gauge fields should be expected to play an important dynamical role. Since they lead to repulsive interactions between the composite fermions, they will disfavor superconducting order \cite{Bonesteel1999} and possibly lead to phase separation \cite{Parameswaran2011}. However, we will instead adopt a more phenomenological approach, analogous to that used in the continuum \cite{Read2000}, in which we simply take the uniform statistical gauge field flux as a fixed background and look for paired states on top of it. 
Our reasons for this are twofold. First, as in the continuum, our motivation is to look for potentially interesting pairing instabilities, not investigate dynamical questions of the stability of these states to gauge field fluctuations. Second, as noted in a previous study \cite{Sohal2018}, mean-field approximations of this type of lattice Chern-Simons action appear to be ``too classical", in the sense that, although the mapping to composite fermions is an exact one, the choice of flux attachment breaks the lattice point-group symmetries. This makes itself manifest in mean-field solutions and, in Ref. \cite{Sohal2018}, the authors do not find uniform density FCI states in their model for this reason \footnote{Theories of this type do not have small expansion parameters. A more correct treatment of quantum fluctuations can lead, presumably, to the partial melting of some of the broken symmetry states found in Ref. \cite{Sohal2018}. The same caveats apply to the problem under consideration here.}.  In the present problem, we are generally not able to find solutions with reasonably small unit cells, if we perform this fully self-consistent analysis. This may be indicative of a similar issue, or of the possibility that we are already seeing the effects of phase separation. In either case, this misses the main physics we which to address which, to reiterate, is the existence of interesting instabilities of the composite fermions.

It should also be noted that we have chosen specific channels into which to decompose the attractive and repulsive interactions. In a fully self-consistent variational calculation, it would be more appropriate to decompose both interactions into all possible channels since, as we shall see, the mean-field solutions typically exhibit CDWs and bond order waves (BOWs), even for $g=0$ \footnote{Here we use a traditional (but inaccurate) terminology in which a CDW is meant a charge density wave on the sites of the lattice, and by a BOW charge density wave on the bonds of the lattice. In the case of an incommensurate CDW, the charge modulation has components both on sites and bonds.}. We have adopted this simplified approach as our goal is not to provide a detailed, quantitative understanding of the phase diagram, but rather to highlight the qualitative features of the phases which may appear in these lattice systems.

With these assumptions and caveats out of the way, we are left with solving for mean-field configurations of spinless fermions in a uniform background magnetic field on a lattice, as described by the mean-field Hamiltonian
\begin{align}
	\notag H_F = &\sum_{\bm x , j} \left[ -t f_{\bm x}^\dg f_{\bm x + \bm e_j}^{\pdg} e^{-ia_{*,j}(\bm x)} + \Delta_{\bm x,j} f^\dg_{\bm x} f^\dg_{\bm x + \bm e_j} + \textrm{H.c.} \right] \\
	& -\mu \sum_{\bm x} f_{\bm x}^\dg f_{\bm x}^{\pdg} + g \sum_{\bm x, \bm y}  f_{\bm x}^\dg f_{\bm x}^{\pdg} U(\bm x - \bm y) \rho(\bm y),
\end{align}
where we have defined $\bm a_* = \bm A - \bm a = (0, \phi_* x)$. We must look for solutions of the following self-consistent equations,
\begin{align}
\rho(\bm x) &= \la f^\dg (\bm x) f (\bm x) \ra \label{eqn:self-consistent-rho} \\
\Delta_{\bm x,j} &= \la f(\bm x+\bm e_j) f(\bm x) \ra \label{eqn:self-consistent-delta}  \\
\sum_{\bm x} \rho(\bm x) &= N_f, \label{eqn:self-consistent-mu}
\end{align}
where $N_f$ is the total number of fermions. For non-zero values of the pairing amplitudes, $\Delta_{{\bm x},j}$, the total fermion number is not conserved by the mean-field Hamiltonian, and so we fix the average density, $n$, by tuning the chemical potential, $\mu$. 

\begin{figure}
  \includegraphics[width=0.31\textwidth]{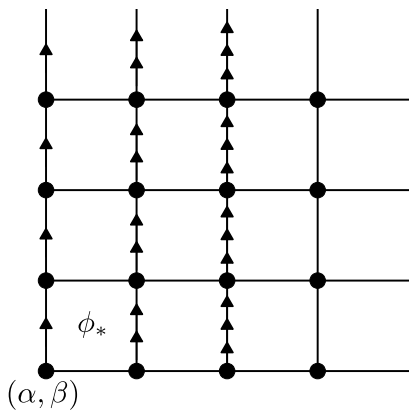}
  \caption{Unit cell used in the mean-field analysis. The net flux per unit cell is $\phi_*$ out of the page. Here we take $\phi_* = 2\pi \left(\frac{5}{8} - \frac{3}{8}\right) = 2\pi \frac{1}{4}$ so that the unit cell contains $q_*\times q_* = 4\times 4$ lattice sites. The arrows represent our choice of the Landau gauge, with the net mean-field gauge field taking the form $\bm a_* = (0, \phi_* \alpha)$. Lastly, $(\alpha,\beta)$ represent the horizontal and vertical coordinates of the lattice sites within a unit cell.}
  \label{fig:unitcell}
\end{figure}

As noted in the previous section, we must allow for pair fields with COM momenta $\bm Q_l + \bm Q_m$, the smallest of which is $(0, 2\pi / q_*)$ and corresponds to a period of $q_*$ lattice sites in the $y$-direction. As such, we will take our unit cell to contain $q_* \times q_*$ lattice sites, as depicted in Fig. \ref{fig:unitcell} for $q_*=4$. This leaves us with $q_*^2$ densities, $\rho_{(\alpha, \beta)}$, and $2q_*^2$ pair fields, $\Delta_{(\alpha,\beta), j}$, to solve for, where $\alpha,\beta = 1 , \dots , q_*$ denote the horizontal and vertical coordinates of the sites within a unit cell (see Fig. \ref{fig:unitcell}). For given values of $V$ and $g$, we numerically solve the saddle-point equations (\ref{eqn:self-consistent-rho})-(\ref{eqn:self-consistent-mu}), using several random \textit{ans\"atze} for the densities and pair fields to ensure we identify the lowest energy solution. Note that the ground state is the solution which minimizes the energy -- not the grand potential -- since we are working at fixed particle number rather than fixed chemical potential. So, although we compute observables within the grand canonical ensemble, we must subtract $-\mu N_f$ from the mean-field Hamiltonian when comparing the energies of different mean-field configurations:
\begin{align}
E = &\la H_F \ra + \mu N_f- \frac{1}{V} \sum_{\bm x , j} |\Delta_{\bm x , j}|^2\nonumber\\
& - \frac{g}{2}\sum_{ \bm x, \bm y } \rho(\bm x) U(\bm x - \bm y) \rho(\bm y).
\end{align}
In the following, we will map out the mean-field phase diagrams as functions of $V$ and $g$.

\subsection{Role of Magnetic Translation Symmetry}

As a brief interlude, let us investigate the role of the magnetic translation symmetry in determining the form of the pair fields \cite{Zhai2010}. In the Landau gauge we have chosen, the magnetic translation operators are given by
\begin{align}
\tilde{T}_1 = \exp\left(i\phi_* \sum_{\bm r} r_2 f^\dg_{\bm r} f^{\pdg}_{\bm r} \right) T_1, \qquad \tilde{T}_2 = T_2, \label{eqn:magnetic-translation-operators}
\end{align}
where $T_{1,2}$ are the ordinary translation operators and have the action $T_j^{-1} f_x T_j = f_{x-e_j}$. The magnetic translations $\tilde{T}_{1,2}$ commute with the kinetic part of the mean-field Hamiltonian. Under the action of $\tilde{T}_1$, the pair fields transform as
\begin{align}
\begin{split}
\tilde{T}_1 \Delta_{(\alpha,\beta),x} \tilde{T}_1^{-1} &= \Delta_{(\alpha+1,\beta),x} e^{-2i\phi_* \beta}, \\
\tilde{T}_1 \Delta_{(\alpha,\beta),y} \tilde{T}_1^{-1} &= \Delta_{(\alpha+1,\beta),y} e^{-2i\phi_* \beta} e^{-i\phi_*}.
\end{split} \label{eqn:mts-pair-field-transformation}
\end{align}
This implies that a mean-field state, $\ket{\psi}$, with, for instance, uniform $p_x+ip_y$ pairing actually breaks magnetic translations and the state $\tilde{T}_1 \ket{\psi}$ will have spatially modulated pair fields.
We alert the reader to this fact now so it is clear, when we present real-space configurations of specific mean-field solutions, that the pair fields of the translated (and rotated) solutions will not take the same form. This is a consequence of the fact that magnetic translations (rotations) are translations (rotations) combined with a gauge transformation and the pair fields are not gauge-invariant quantities.

Let us now consider solutions which preserve the magnetic translation symmetry, so that $\tilde{T}_j \Delta_{(\alpha,\beta),j} \tilde{T}_j^{-1} = \Delta_{(\alpha,\beta),j}$. On defining the Fourier transform of the pair fields in the $y$-direction,
$\Delta_{\alpha,P_l,j} = \sum_{\beta=1}^{q_*} \Delta_{(\alpha,\beta),j} e^{-iP_l \beta}$ with $P_l = \frac{2\pi l}{q_*}, \, l \in \mathbb{Z},$
and imposing the above magnetic translation symmetry constraints, we find
\begin{align}
\Delta_{\alpha,P_l,j} = \Delta_{\alpha+1,P_l+2\phi_*,j} e^{-i\phi_* \delta_{j,y}}. \label{eqn:mts-momentum-constraint}
\end{align}
This implies that zero-momentum pairing will generically coexist with finite-momentum pairing, if magnetic translation symmetry is preserved. Of course, there is no guarantee that magnetic translations will be respected by the mean-field ground state and we will often find it to be the case that it is not. Nevertheless, this observation highlights the point that there is a predisposition to finite-momentum pairing in these lattice systems.

\section{Fermionic Paired FCI Phase Diagrams}

\begin{figure*}
  \includegraphics[width=0.95\textwidth]{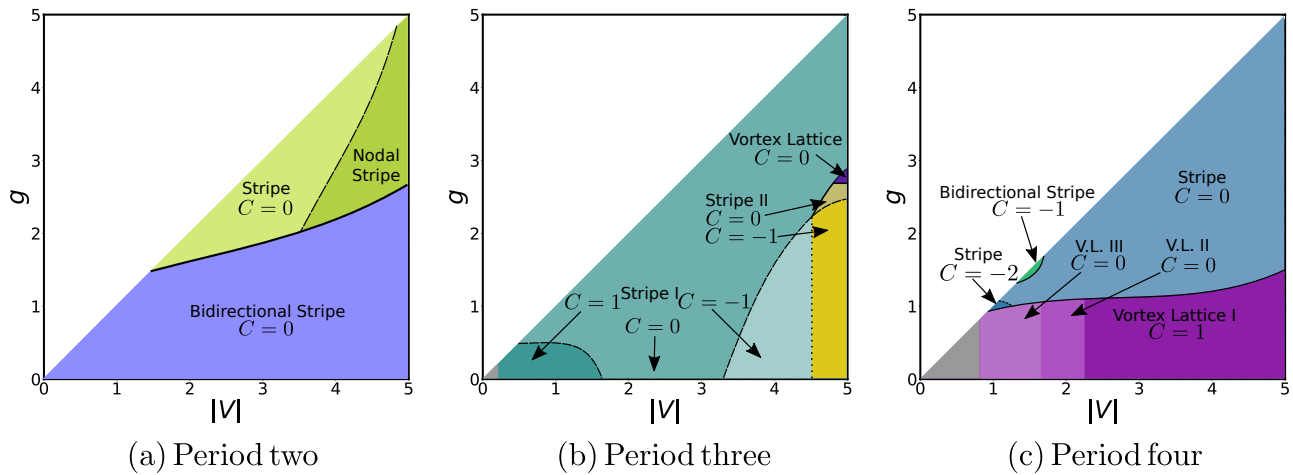}
  \caption{Schematic mean-field phase diagrams as functions of the NN attraction, $|V|=-V$, and NNN repulsion, $g$, for the fermionic configurations of Table \ref{tab:fillings}. Solid (dashed) black lines correspond to first order (continuous) transitions. The dotted line separating the Stripe I and II regions in (b) indicates a crossover. Gapped phases are labeled by the Chern number, $C$, of the BdG bands. The grey regions indicate where the energies of the saddle-point equation solutions are too close to numerically deduce which is the ground state. Details of the phases are presented in the main text and illustrated in Figures \ref{fig:period-2-mf-configs}, \ref{fig:period-3-mf-configs}, and \ref{fig:period-4-mf-configs}.}
  \label{fig:Phase-Diagrams}
\end{figure*}

The results of our self-consistent mean-field analysis are summarized in the phase diagrams of Fig. \ref{fig:Phase-Diagrams}. We find a host of translation symmetry breaking paired states of the composite fermions, the qualitative features of which we now describe in more detail. In addition to the site-centered charge density and pair field configurations, we characterize these phases by computing the link currents and the bond densities,
\begin{align}
	\la j_{\bm x , k} \ra &= \la i f^\dg_{\bm x}  f_{\bm x+ \bm e_k} e^{-i a_{*,k}(\bm x,t)} + \textrm{H.c} \ra , \\
	\la B_{\bm x, k} \ra &= \la f^\dg_{\bm x}  f_{\bm x+ \bm e_k} e^{-i a_{*,k}(\bm x,t)} + \textrm{H.c} \ra ,
\end{align}
as well as the Chern number, $C$, of the Bogoliubov-de Gennes (BdG) band structure, using the method of Ref. \cite{Fukui2005}. The latter quantity determines the number and chirality of Majorana edge modes in a system with open boundary conditions. This allows us to determine the topological order of the system via the bulk-boundary correspondence, on taking into account the presence of a charged chiral boson from the gapped charge sector. Equivalently, from the bulk perspective, vortices of the pair field will trap $C$ Majorana zero modes (MZMs). 

Much as in the well studied case of the paired FQH states in the continuum \cite{Read2000}, due to the Higgsing of the dynamical Chern-Simons gauge field by the pairing amplitudes, vortices of the pair field are finite energy excitations and carry a charge $e/4k$, where $e$ is the charge of the electron. So, states with an odd Chern number possess non-Abelian topological order, as these pair field vortices will possess one unpaired MZM. Conversely, states with an even Chern number possess Abelian topological order. In particular, since we have focused on FQH states arising from attaching a single pair of flux quanta $(k=1)$, states with  $C=1,-1$ possess the same topological order as the Pfaffian and PH-Pfaffian states \cite{Son2015}, respectively, whereas those with $C=0$ support the same topological order as the Abelian Halperin paired state \cite{Halperin1983}. 

The relation between the Higgsing of the Chern-Simons gauge field and the non-Abelian topological order is a subtle issue. Its root reason is the the fact that the pair field condensate leaves a local $\mathbb{Z}_2$ symmetry unbroken in a regime in which the theory is deconfined \cite{Fradkin-1979}. An example is the case of a conventional superconductor coupled to a dynamical gauge field which has $\mathbb{Z}_2$ topological order \cite{Hansson-2004}. In the case of a relativistic field theory, the non-Abelian character can be described either through a similar pairing mechanism, or in terms of a topological phase of the partition function in the form of an $\eta$-invariant \cite{Seiberg2016a}.

\subsection{Period Two}

\begin{figure}
  \includegraphics[width=0.45\textwidth]{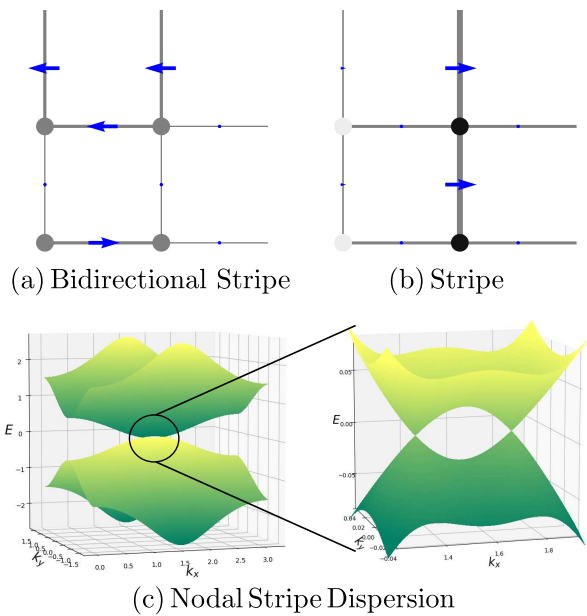}
   \caption{(a) and (b): Period-two mean-field configurations. In this and the following figures, the color of the sites indicates the charge density, with darker (lighter) sites corresponding to higher (lower) density. Likewise, the width of the links represent the magnitude of the bond density, $B_{\bm x, j}$. The blue arrows represent the pair fields $\Delta_{\bm x , j} = |\Delta_{\bm x , j}|e^{i\theta_{\bm x , j}}$, with length proportional to $|\Delta_{\bm x , j}|$ and angle relative to the horizontal given by $\theta_{\bm x , j}$. The link currents all vanish. (c): Spectrum of the BdG Hamiltonian for mean-field configuration (b). The left panel depicts the two bands closest to $E=0$. The black circle highlights the presence of two Majorana cones along the line $k_y = 0$, which are depicted in more detail in the right panel.} \label{fig:period-2-mf-configs} 
\end{figure}

We now turn to the non-uniform paired phases. We begin with the period-two phase diagram, depicted in Fig. \ref{fig:Phase-Diagrams}a, in which we find three striped phases. The real space configurations of these phases are depicted in Fig. \ref{fig:period-2-mf-configs}. We note that the net (statistical plus magnetic) flux per plaquette is $\pi$ and so, prior to the addition of interactions, the mean-field composite Fermi liquid solution (Fig. \ref{fig:FermiSurface}a) preserves time reversal symmetry (TRS). The mean-field paired ground states we find also preserve TRS since all the pair fields, $\Delta_{\bm x , j}$, can be made real by a global $U(1)$ rotation. 

Focusing on the individual phases in more detail, the ground state for small $g$ is a bidirectional stripe phase. As depicted in Fig. \ref{fig:period-2-mf-configs}a, this state possesses a uniform site density but also a bidirectional BOW. In particular, $B_{\bm x , x}$ and $B_{\bm x , y}$ possess modulations at the wave vectors $(\pi , 0)$ and $(0 , \pi)$, respectively. This is not surprising, as the pair fields take the forms $\Delta_{\bm x , x} = \Delta e^{i\bm q_1 \cdot \bm x} + \tilde{\Delta} e^{i\bm q_2 \cdot \bm x}$ and $\Delta_{\bm x , y} = \tilde{\Delta} e^{i\bm q_1 \cdot \bm x} - \Delta e^{i\bm q_3 \cdot \bm x}$, where $\tilde{\Delta} > \Delta > 0$, $\bm q_1 = (0,\pi)$, $\bm q_2 = (\pi,\pi)$ and $\bm q_3 =(0,0)$.
In general, the presence of pair fields at momenta $\bm q_1$ and $\bm q_2$ will induce a \emph{daughter} CDW order with amplitude $\rho_{\bm q_1 - \bm q_2} \sim \Delta_{\bm q_1} \Delta_{\bm q_2}^*$ $+ \Delta_{-\bm q_2} \Delta_{-\bm q_1}^*$, where $\rho_{\bm q}$ is the Fourier transform of the charge density, as can be shown through a simple free energy analysis \cite{Agterberg2008,Berg2009}. However, in the present problem, we must be careful to note that the phases of the pair fields, and hence their Fourier components, depend on the choice of gauge for the background flux. In particular, as noted above, the pair fields transform non-trivially under magnetic translations and rotations. As such, we cannot directly use the free energy analysis of Ref. \cite{Agterberg2008} to deduce the daughter orders of the spatially modulated superconducting order. A more careful treatment, which is beyond the scope of the present work, would require the analysis of a free energy which takes into account the transformations of the pair fields under the magnetic algebra. Nevertheless, it is clear that we can still identify the BOW as a daughter order of the striped superconducting order (and hence a consequence of finite momentum pairing of the composite fermions) by virtue of the fact that this phase exists as the ground state in the absence of the NNN repulsive interaction, at $g=0$.

The band structure of the BdG Hamiltonian for this phase is less interesting. It is fully gapped with $C=0$, implying there are no chiral Majorana edge states. We have also studied this mean-field configuration with open boundary conditions to confirm that there are indeed no edge states protected by the mean-field TRS or any other symmetry.

As $g$ is increased, there is a first-order transition to a striped $p_y$ phase, in which $\Delta_{\bm x , x} = 0$, while $\Delta_{\bm x , y} = \tilde{\Delta} + \Delta e^{i \bm q \cdot \bm x}$ with $\bm q = (\pi, 0)$ and $\tilde{\Delta} > \Delta > 0$. The modulation of the pair fields in this phase appears to be driven by the $(\pi, 0)$ CDW engendered by the repulsive NNN interactions, as this phase does not exist as a solution of the saddle-point equations at $g=0$. Moreover, we have numerically checked that a similar stripe phase can be obtained in a square lattice system with the same interactions, but with a vanishing magnetic flux and hence a single Fermi pocket. 
Nevertheless, the BdG spectrum exhibits an interesting nodal structure. For large $g$ and $V$, the system possesses two Majorana cones, as shown in Fig. \ref{fig:period-2-mf-configs}c. As $g$ is increased further or $V$ decreased, the cones approach and annihilate one another (indicated by the dashed black line in Fig. \ref{fig:period-2-mf-configs}), yielding a fully gapped spectrum.  

\begin{figure}
  \includegraphics[width=0.48\textwidth]{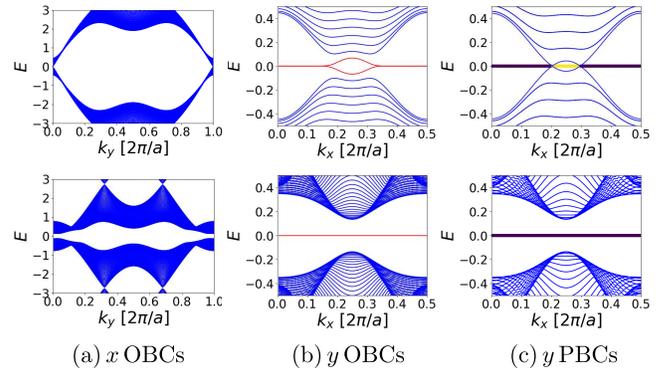}
  \caption{Dispersions of the (top row) nodal and (bottom row) gapped stripe phases on finite-size systems with different boundary conditions. In (c), we also plot a horizontal line at $E=0$, representing the topological invariant $\mathcal{M}(k_x)$ defined in Appendix \ref{sec:appendix-majorana-flat-band}. Purple (yellow) indicates $\mathcal{M}(k_x)=-1(+1)$.}
  \label{fig:majorana-flatband-dispersions}
\end{figure}

Although $C=0$ in the gapped stripe phase, both the nodal and gapped phase band structures are in fact topologically non-trivial. This is demonstrated in Fig. \ref{fig:majorana-flatband-dispersions}, in which we plot the energy spectra for these phases on finite-size systems with open and periodic boundary conditions (OBCs and PBCs, respectively). In the nodal phase, on imposing OBCs along the direction parallel to the stripes, we find a Majorana flat band connecting the projections of the bulk nodes onto the edge Brillouin zone (BZ). In the gapped phase, we find a Majorana flat band spanning the entire surface BZ. These properties are typical of $p_x$-paired states \cite{Sato2011}. We show in Appendix \ref{sec:appendix-majorana-flat-band} that a combination of the particle-hole symmetry of the BdG Hamiltonian and reflection symmetry, with the reflection axis taken along a stripe, are sufficient to protect these flat bands and the nodal points. 

Physically, the existence of these flat bands is not surprising, as the mean-field ground state resembles an array of Kitaev chains \cite{Kitaev2001}. At large values of $g$, hopping between the chains consisting of sites with high density, which also have non-zero $\Delta_{\bm x , y}$, will be suppressed due to the intervening low density chains and the NNN repulsion. This yields an array of decoupled Kitaev chains which, in the topological regime, will host MZMs at their ends when OBCs are imposed, giving rise to the observed Majorana flat band.

Since they both have $C=0$, the bidirectional stripe phase and gapped stripe phases possess the topological order of the Abelian Halperin paired state. That being said, based on the physical picture of the gapped stripe phase as an array of nearly decoupled Kitaev chains, we expect that lattice dislocations should bind MZMs (see Appendix \ref{sec:appendix-dislocations}). This is a particularly intriguing possibility in the context of cold atom experiments, where lattice defects can be engineered directly. A somewhat similar nematic FQH phase was found in a coupled wire construction of paired FQH states in Ref. \cite{Kane2017}, although, in that case, the edge supported a pair of helical Majoranas with finite dispersion. Lastly, we note that the nodal striped phase is a quantum Hall thermal semi-metal in that charged excitations are gapped in the bulk, but the gapless Majoranas can still transport heat. The nodal striped phase is not strictly topologically ordered since it has a gapless spectrum. Nevertheless, it still supports gapped charge-$1/2$ Laughlin quasiparticles.

\subsection{Period Three}
\begin{figure}
    \includegraphics[width=0.48\textwidth]{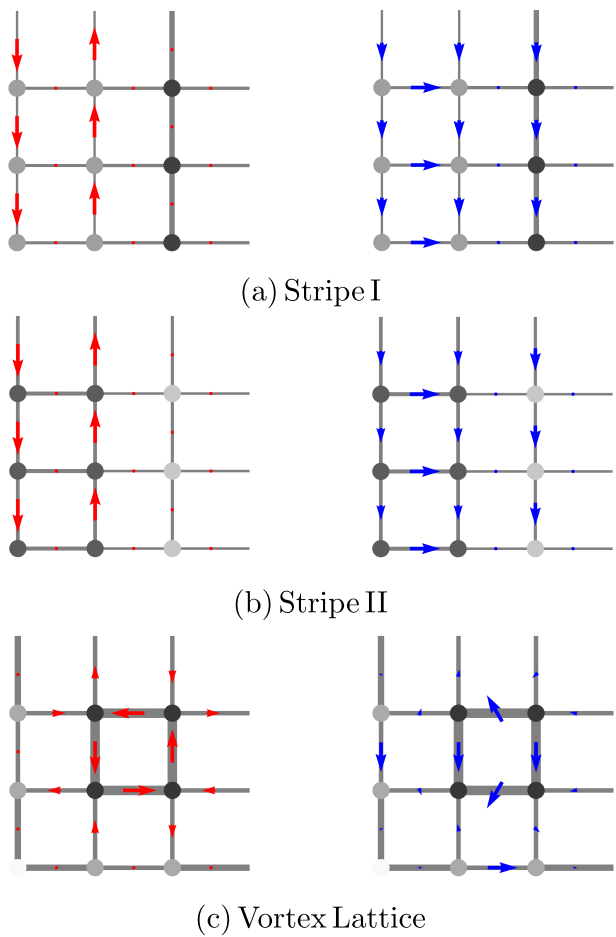}
  \caption{Period-three mean-field configurations. For each configuration, the left figure depicts the link currents, $j_k$, as red arrows, while the right figure depicts the pair fields in the same manner as Fig. \ref{fig:period-2-mf-configs}.} \label{fig:period-3-mf-configs}
\end{figure}

We will now discuss the period-three inhomogeneous paired states. As shown in Fig. \ref{fig:Phase-Diagrams}b, the period-three phase diagram is dominated by unidirectional stripe phases. The real space configurations of these phases are depicted in Fig. \ref{fig:period-3-mf-configs}. The stripe I and II configurations clearly belong to the same phase -- they both possess a CDW at wave vector $(2\pi / 3 , 0)$. 
For small $g$, as $|V|$ is increased, there is a crossover from stripe I to stripe II, as the CDW order parameter continuously drops to zero around $|V| \approx 4.5$ and then changes sign. This crossover is indicated by the dotted black line in Fig. \ref{fig:Phase-Diagrams}b. As $g$ is increased, however, this crossover changes to a first order transition around $g \approx 2.5$. The stripe I/II configurations are also characterized by finite momentum pairing and counter-propagating currents.
The pair fields, in the chosen gauge, have the forms $\Delta_{\bm x , x} = \Delta_0 + \Delta_{1} \cos(2\pi x / 3)$, where $\Delta_0>\Delta_1 >0$, and $\Delta_{\bm x , y} = -i\Delta_2 - i \Delta_3\cos(2\pi(x+1)/3)$, where $\Delta_3>\Delta_4>0$. Note that the pair fields on the horizontal links of the rightmost two columns in Figures \ref{fig:period-3-mf-configs}a and \ref{fig:period-3-mf-configs}b do not vanish; they are simply about one to two orders of magnitude smaller than the pair fields on the other links. As in the example of the period-two bidirectional stripe, we identify the CDW order as a daughter order of the striped superconducting order, by virtue of the fact that the the stripe I/II phases persist down to $g=0$. 
Aside from the stripe I/II phases, there is a small region of the phase diagram around $(V,g)=(-5,3)$ which supports a vortex lattice phase and is separated from the other phases by a first order transition. As shown in Fig. \ref{fig:period-3-mf-configs}c, this phase consists of an array of clusters of four high density sites, around which there are circulating currents. As for the topological properties of these states, the stripe I/II phase supports regions with $C=-1,0,1$ and $C=0,-1$, respectively, whereas the vortex lattice phase has Chern number $C=0$. So, in contrast to the period-two case, non-Abelian phases with the topological orders of the Pfaffian and PH-Pfaffian are present in the period-three phase diagram.

We note that that we are not able to conclusively identify the ground state in the unlabeled grey region of Fig. \ref{fig:Phase-Diagrams}b. Here, several states (with $C=\pm 1$) compete with the stripe I configuration and all are nearly degenerate, up to our chosen numerical precision. This suggests that the system will likely be unstable to phase separation in this regime.

\subsection{Period Four}

\begin{figure}
    \includegraphics[width=0.48\textwidth]{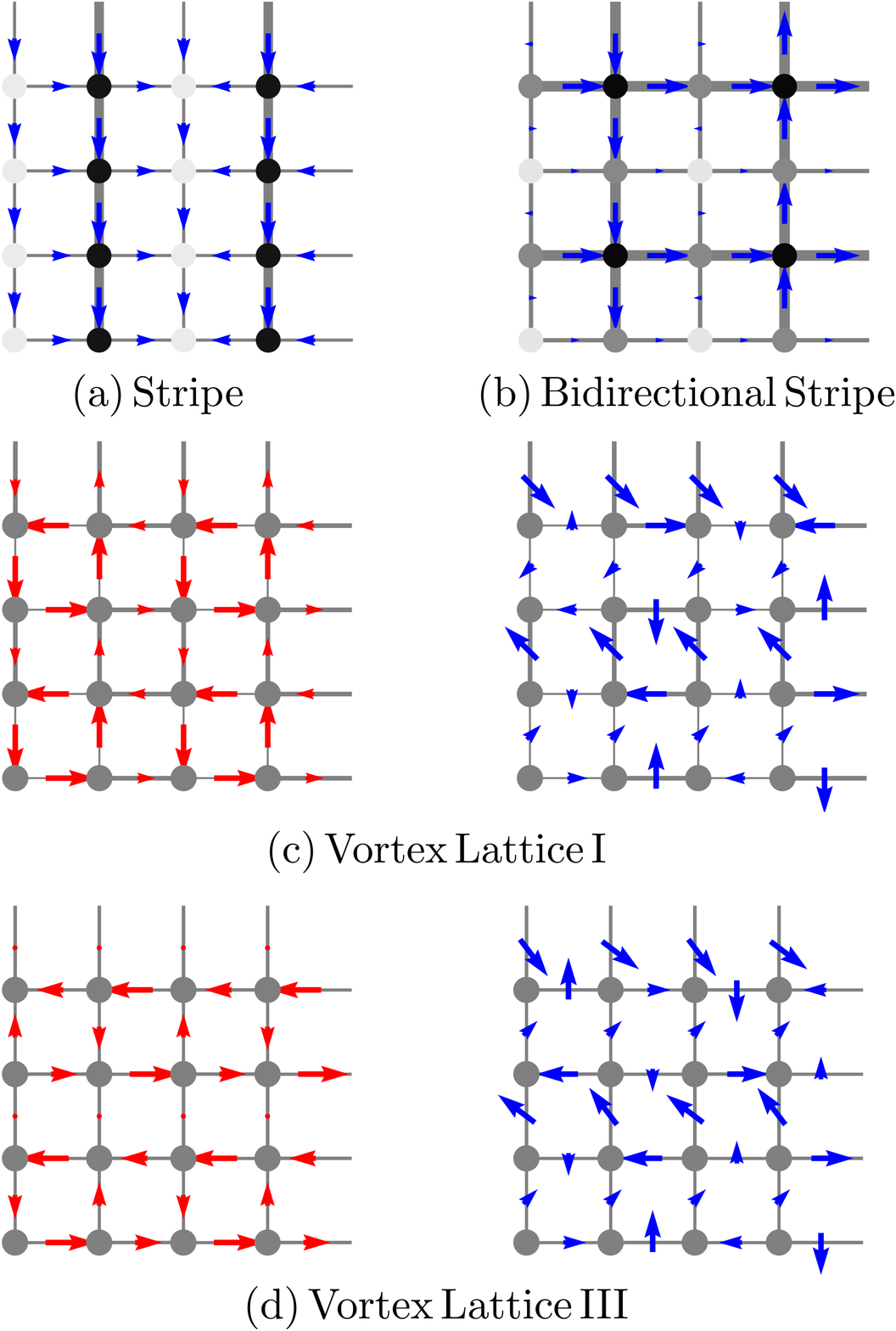}
  \caption{Period-four mean-field configurations. The link currents in the stripe configurations (a,b) all vanish.} \label{fig:period-4-mf-configs}
\end{figure}

Lastly, we have the period-four phase diagram, shown in Fig. \ref{fig:Phase-Diagrams}c, which exhibits the greatest diversity of phases. A unidirectional stripe phase, shown in Fig. \ref{fig:period-4-mf-configs}a, occupies most of the the $g \gtrapprox 1$ region. Of note is the fact that it supports a CDW and a BOW in the $x$-direction with a period of two lattice sites, while the pair field modulation has a period of four sites in the same direction (that is, the pair field pattern returns to itself after four \emph{magnetic} translations). 
Explicitly, the pair fields on the $y$-links, $\Delta_{\bm x , y}$, possess a uniform $\bm q_1 = (0,0)$ Fourier component and a modulation at wave vector $\bm q_2 = (\pi , 0)$, while the pair fields on the $x$-links are given by $\Delta_{\bm x , x} = \Delta_{\bm q}e^{i\bm q \cdot \bm x} + \Delta_{-\bm q} e^{-i \bm q \cdot \bm x}$, where $\bm q = (\pi / 2, 0)$ and $\Delta_{\bm q} = |\Delta|e^{-i\pi / 4} = \Delta_{-\bm q}^*$. 
Note that the appearance of a CDW with half of the period of the pair field modulation is characteristic of PDW states \cite{Agterberg2008}; indeed, the form of $\Delta_{\bm x , x}$ is precisely that of a PDW, at least in the chosen gauge. In fact, this unidirectional stripe phase remains a solution of the saddle-point equations down to $g=0$, and so it indeed owes its existence to finite-momentum pairing of the CFs -- the NNN repulsive interactions are needed only to stabilize it as the ground state. Additionally, it is topologically trivial except for a small region of the phase diagram around $(V,g) = (-1.2,1.2)$, where the BdG bands have $C=-2$. In this regime, the edge of the system supports a chiral boson from the charge sector and two counter-propagating Majorana fermions. 

The NNN repulsion also helps stabilize a bidirectional stripe phase, shown in Fig. \ref{fig:period-4-mf-configs}b, in a small region of the phase diagram. This phase possesses CDWs at wave vectors $(0,\pi)$,  $(\pi, 0)$, and $(\pi,\pi)$ as well as modulations of the bond densities, $B_{\bm x , x}$ and $B_{\bm x , y}$, at wave vectors $(0,\pi)$ and $(\pi, 0)$, respectively. The pair fields take the form
\begin{align}
	\Delta_{\bm x , x} &= \tilde{\Delta} + \Delta + (\tilde{\Delta} - \Delta)e^{i\pi y} \\
	\Delta_{\bm x , y} &= i\tilde{\Delta}\cos(\pi x / 2 + \pi / 2) + \Delta e^{i\pi y} \cos(\pi x / 2),
\end{align}
with $\tilde{\Delta} > \Delta > 0$. Note that this mean-field configuration is invariant under two \emph{magnetic} translations along both lattice directions, and so the CDW and BOW has the same periodicity as the pair field modulation, in contrast to the unidirectional stripe phase. However, we find that this phase exists as a (metastable) self-consistent mean-field solution at $g=0$, and so it seems reasonable to view the CDWs and BOWs as daughter orders of the spatially modulated pairing.
As far as its topological properties are concerned, this bidirectional stripe phase has Chern number $C=1$, and so possesses the topological order of the Pfaffian.

In the region below $g \approx 1$, we find competition between various configurations with circulating currents, which we refer to as vortex lattices. Two examples of these phases are shown in Fig. \ref{fig:period-4-mf-configs}c,d. For $|V| > 2.2$, the ground state is the vortex lattice I phase, which exhibits a square lattice of vortices. It also has $C=-1$, and so supports the same topological order as the PH-Pfaffian. As $|V|$ is lowered, the system transitions through other vortex lattice phases, including that of Fig. \ref{fig:period-4-mf-configs}d, in which there appears to be a triangular lattice of vortices. In the region $|V| \lessapprox 0.8$, marked by the color gray in Fig. \ref{fig:Phase-Diagrams}c, we find competition between several vortex lattice states, one of which has Chern number $C=-2$. These solutions appear to be degenerate (up to numerical precision), suggesting the system will likely be unstable to phase separation. It is thus unclear whether a uniform paired state of CFs can actually be stabilized in this regime, or if a proliferation of vortices will return the system to a composite Fermi liquid.

\section{Bosonic Paired FCI Phase Diagrams \label{sec:appendix-bosonic-states}}

\begin{table}
\caption{\label{tab:boson-fillings}%
Details of the three composite Fermi liquid states for the bosonic system. Here, $2k'-1$ is the number of attached statistical flux quanta.
}
\begin{ruledtabular}
\begin{tabular}{ccccccc}
 &
$\phi_0/2\pi$&
$n$ &
$\nu$ &
$k'$ &
$\phi/2\pi$ &
$\phi_*/2\pi$\\
\colrule
Period two & $3/4$ & $1/4$ & $1/3$ & $1$ & $1/4$ & $1/2$ \\
Period three & $2/3$ & $1/9$ & $1/6$ & $2$ & $1/3$ & $1/3$ \\
Period four & $5/8$ & $1/8$ & $1/5$ & $2$ & $3/8$ & $1/4$
\end{tabular}
\end{ruledtabular}
\end{table}

\begin{figure}
  \includegraphics[width=0.48\textwidth]{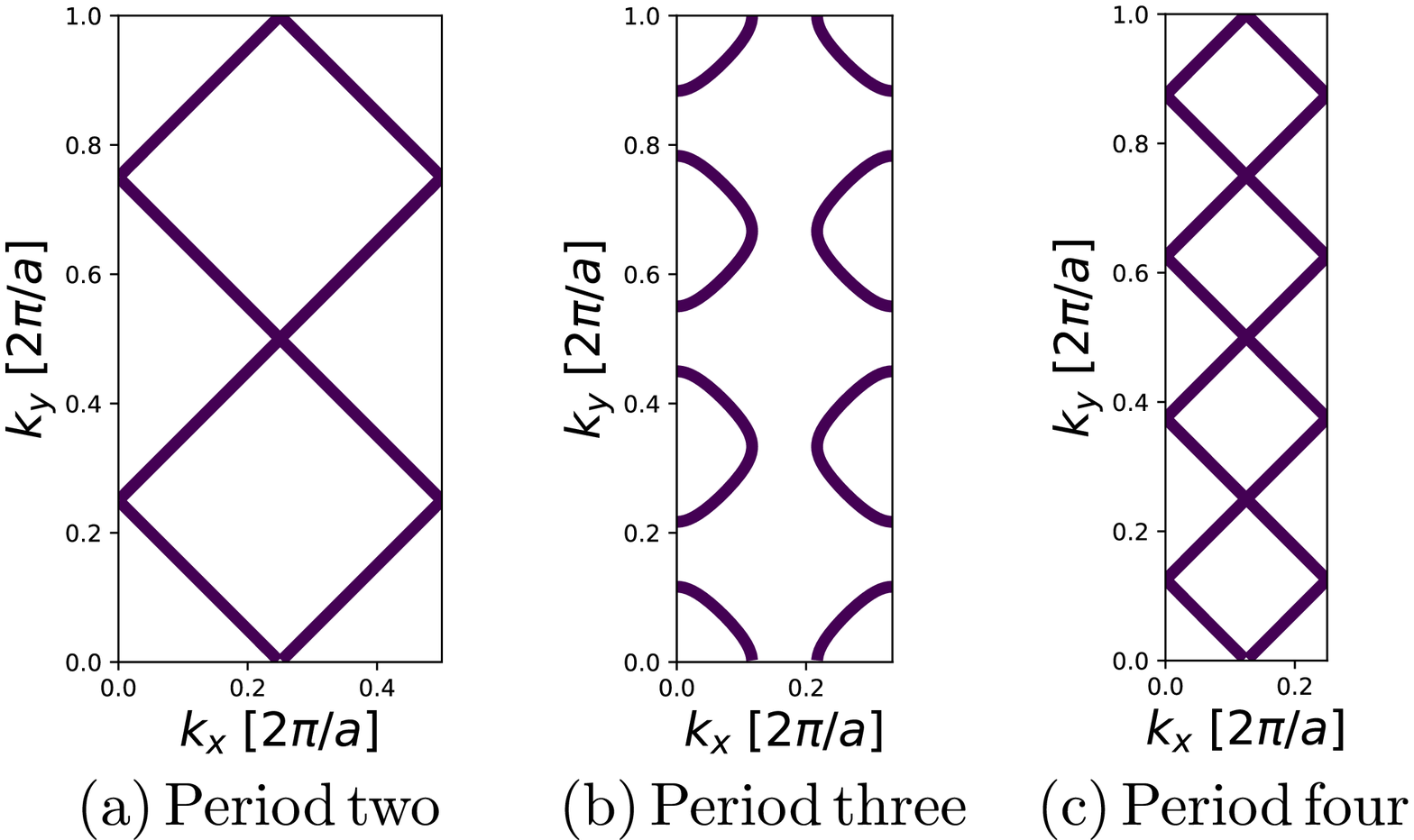}
  \caption{Composite Fermi surfaces for the period two, three, and four configurations given in Table \ref{tab:boson-fillings}.}
  \label{fig:boson-FermiSurface}
\end{figure}

\begin{figure*}
  \includegraphics[width=0.95\textwidth]{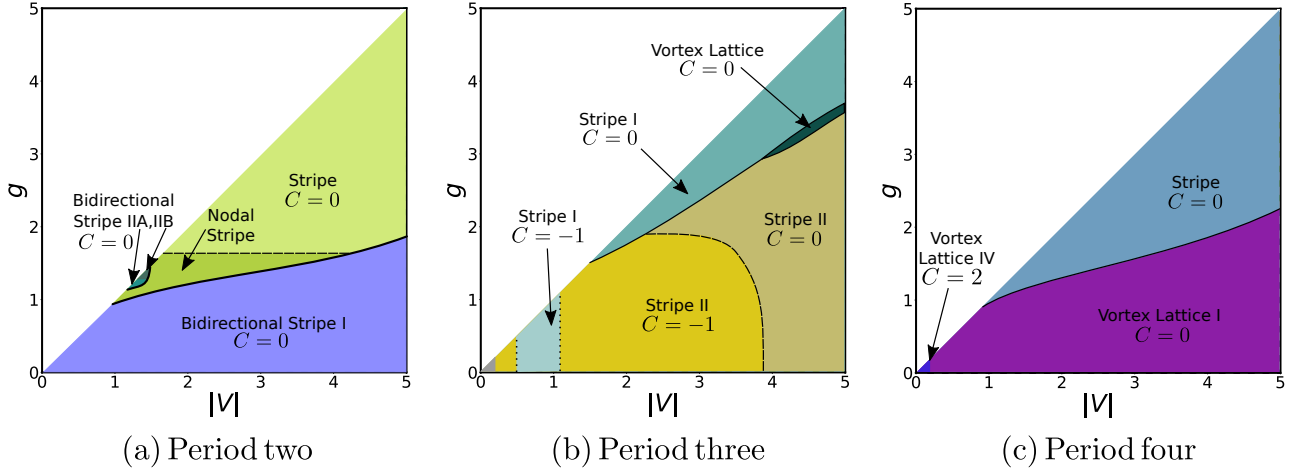}
  \caption{Schematic mean-field phase diagrams as functions of the NN attraction, $|V|=-V$, and NNN repulsion, $g$ for the bosonic configurations listed in Table \ref{tab:boson-fillings}. The bidirectional stripe I phase in (a) is the same as the bidirectional stripe phase present in Fig. \ref{fig:Phase-Diagrams}a.}
  \label{fig:boson-Phase-Diagrams}
\end{figure*}

\begin{figure}
  \includegraphics[width=0.48\textwidth]{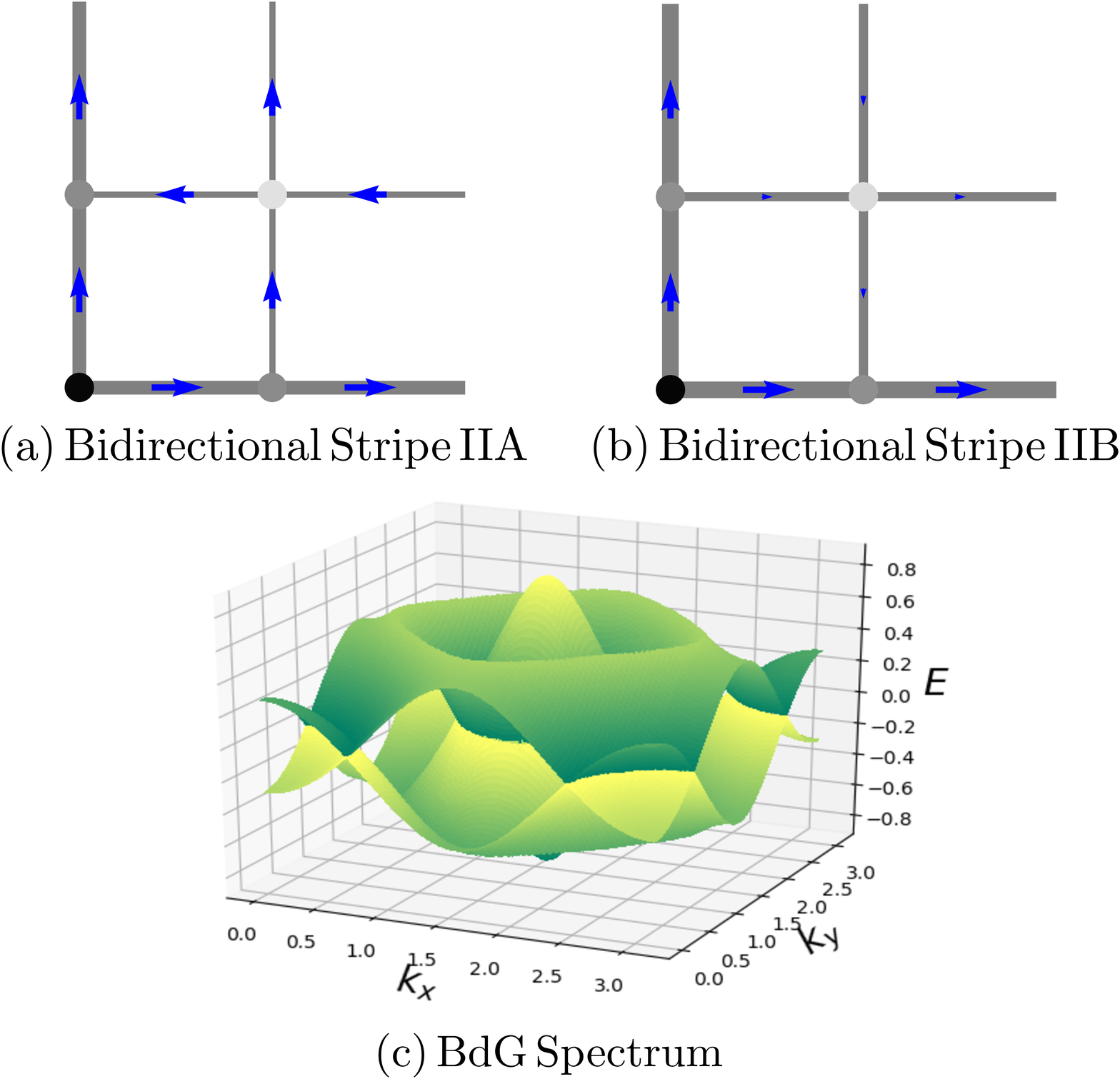}
   \caption{Real-space configuration of the bidirectional stripe (a) IIA and (b) IIB phases in the bosonic FCI period-two phase diagram (Fig \ref{fig:boson-Phase-Diagrams}a). The link currents vanish in both configurations. (c) The two BdG bands closest to $E=0$ for IIB (the spectrum for IIA is similar). Despite appearances, there is a very small gap, as the pair fields are non-zero but small.} \label{fig:boson-period-2-mf-configs} 
\end{figure}

Thus far, we have considered paired FCI states in a tight-binding model of fermions. In this section, we repeat our analysis for a system of hardcore bosons in the same square-lattice Hofstadter model, which is of relevance for cold atom experiments \cite{Aidelsburger2013,Miyake2013,Aidelsburger2014}. The set-up is the same as that of the fermionic case considered above, with the only difference being that we must attach an odd number of flux quanta to the bosons to obtain a theory of composite fermions. Hence, we must take the Chern-Simons coupling to be
\begin{align}
	\theta = \frac{1}{2\pi(2k'-1)}, \quad k' \in \mathbb{Z}.
\end{align}
In gapped, paired states of the CFs, vortices of the pair field will thus carry charge $e/(4k'-2)$. We again consider three different configurations of filling and background magnetic flux, as summarized in Table \ref{tab:boson-fillings}, such that the composite fermions form Fermi surfaces with two, three, and four pockets, as shown in Fig. \ref{fig:boson-FermiSurface}.

Repeating the same mean-field analysis as for the fermionic problem, we obtain the phase diagrams of Fig. \ref{fig:boson-Phase-Diagrams}, which exhibit nearly the same topology as the corresponding fermionic phase diagrams. One novel feature is the emergence of the bidirectional stripe IIA and IIB phases in the period-two phase diagram (Fig. \ref{fig:boson-Phase-Diagrams}a) around $(V,g)=(-1.3,1.3)$, which support CDWs at wave vectors $(0,\pi)$, $(\pi , 0)$, and $(\pi , \pi)$ and BOWs in $B_{\bm x , x}$ and $B_{\bm x , y}$ at wave vectors $(0,\pi)$ and $(\pi , 0 )$, respectively. The real space configuration of these phases are depicted in Fig. \ref{fig:boson-period-2-mf-configs}; the differences between IIA and IIB are that, in the former, the $(0,0)$ component of $\Delta_{\bm x , y}$ is greater than the $(\pi,0)$ component and the $(0,\pi)$ component of $\Delta_{\bm x , x}$ is greater than the $(0,0)$ component, while the opposite statements hold true in the latter. Unlike the other period-two phases, these phases spontaneously breaks TRS since the pair fields on the $x$-links, $\Delta_{\bm x , x}$, are all real while those on the $y$-links, $\Delta_{\bm x , y}$, are imaginary. The pair fields have very small magnitudes, yielding a minute gap which is not easily seen in Fig. \ref{fig:boson-period-2-mf-configs}c. As such, even at low temperatures, the system will exhibit unquantized heat transport mediated by the Bogoliubov quasiparticles through the bulk.

Another new phase seems to appear at small $|V|$ and $g$ in the period four diagram; since the order parameters are so small in this region, it is difficult to conclusively identify the nature of this phase, but we tentatively describe it as a vortex lattice and label it as Vortex Lattice IV. The BdG band structure has $C=2$, and so this phase is Abelian. We note that a different $C=2$ paired quantum Hall phase was studied in Ref. \cite{Santos2019}, which resulted from somewhat similar PDW physics.

\section{Discussion and Conclusion}

We have presented a qualitative, mean-field picture of the intertwining of symmetry breaking and topological order in FCI states arising from the finite momentum pairing of composite fermions. This is a consequence of magnetic translation symmetry enforcing the presence of multiple composite Fermi pockets. We find a diverse array of paired states, the most notable of which exhibit some subset of the following observable features:
\begin{enumerate}
	\item Daughter CDW and/or BOW order arising from the modulated pair fields, similar to that in theories of PDW states in the cuprates.
	\item Gapped neutral sectors possessing Chern numbers $C=-2,-1,0,1,2$, resulting in $C$ Majorana edge modes with chirality $\mathrm{sgn} (C)$, in addition to a charged chiral boson. Here, $C=-1,0,1$ correspond to the PH-Pfaffian, Halperin paired state, and Pfaffian topological orders, respectively.
	\item Gapless neutral sectors, forming quantum Hall thermal semimetals (only the nodal stripe phase in Figures \ref{fig:Phase-Diagrams}a and \ref{fig:boson-Phase-Diagrams}a possesses this property).
	\item The possible trapping of MZMs by lattice dislocations (only the gapped stripe phase in Figures \ref{fig:Phase-Diagrams}a and \ref{fig:boson-Phase-Diagrams}a possesses this property).
\end{enumerate}
Although the more interesting phases we find occupy small regions of the phase diagram, there is some hope for observing these states in future cold atom experiments, in which the nature of the interactions can be finely tuned. At a minimum, our results demonstrate that the observation of BSO in an experimental setting need not rule out concomitant TO, as their coexistence is in fact a generic scenario in the composite fermion picture. In particular, the CDW patterns we discuss could be directly imaged in cold atom experiments \cite{Koepsell2020}.

Looking forward, it may prove interesting to better understand the properties of lattice defects in these systems. In particular, we found a stripe phase in the period-four phase diagram exhibiting a CDW with half the period of the pair field modulation, a feature shared by PDWs. In PDW states, a dislocation of the CDW pattern will require the pair field phase to wind by $2\pi$ about the dislocation, trapping a vortex \cite{Agterberg2008}. It is possible that lattice dislocations in this phase, or related paired FCI states, may display similar properties. If the BdG band structure has Chern number $C$, such vortices would trap $C$ MZMs and would provide a novel way of engineering non-Abelian defects in a manner distinct from previous proposals \cite{Barkeshli2012,Liu2017}.

Coupled wire constructions \cite{Sondhi-2001,Kane2002,Teo2014,Kane2017} may also provide a means by which to demonstrate the existence of the striped states we find beyond mean-field, especially since, by definition, these are anisotropic states. However, while such constructions would allow us to identify regions of the phase diagram in which these states could in principle exist through the fine-tuning of interactions, there is still the issue of phase separation. As we have discussed, there appear to be nearly degenerate vortex lattice solutions in the fermionic period-four phase diagram, suggesting a tendency towards a proliferation of vortices and hence a destruction of superconducting order. Although the more interesting stripe phases can be stabilized, as we have seen, through long range repulsive interactions, such phases are also sensitive to the breaking of translation symmetry via, for instance, disorder or the harmonic traps used in cold atom experiments. Such features would result in local variations of the density with periods which may be incommensurate with the expected stripe order in a clean system. So, while cold atoms experiments and solid state Moir\'e systems provide promising platforms in which to search for our proposed finite momentum paired FCI states, there are several physical hurdles which may disfavor the realization of said states.

\begin{acknowledgments}

We acknowledge helpful discussions with I. Bloch and E.W. Huang. R.S. acknowledges the support of the Natural Sciences and Engineering Research Council of Canada (NSERC) [funding reference number 6799-516762-2018]. This work was also supported in part by the US National Science Foundation under grant No. DMR-1725401 at the University of Illinois. This work made use of the Illinois Campus Cluster, a computing resource that is operated by the Illinois Campus Cluster Program (ICCP) in conjunction with the National Center for Supercomputing Applications (NCSA) and which is supported by funds from the University of Illinois at Urbana-Champaign.

\end{acknowledgments}

\appendix

\section{Details of Flux Attachment \label{sec:appendix-fluc-attachment-details}}

We direct the reader to Refs. \cite{Eliezer1992,Sun2015} for more details about the lattice Chern-Simons action of which we make use. Here, we record only for completeness the explicit form of the $\mathcal{K}$-matrix (not to be confused with the $K$-matrix appearing in multi-component Abelian Chern-Simons theories), which does not play a role in our mean-field analysis:
\begin{align}
\mathcal{K} = - \frac{1}{2} \begin{pmatrix}
d_2 \hat{d}_2 && -2 - 2d_1 + 2 \hat{d}_2 + \hat{d}_2 d_1 \\
2 + 2 d_2 - 2 \hat{d}_1 - \hat{d}_1 d_2 && -d_1 - \hat{d}_1
\end{pmatrix}.
\end{align}
The form of $\mathcal{K}$ is lattice dependent and ensures that the theory is gauge invariant. We note that the lattice Chern-Simons action of Ref.  \cite{Sun2015} can only be defined on lattices for which the number of vertices matches the number of plaquettes, of which the square lattice is one example. 

\section{Topological Properties of Period-Two Stripe Phases}
\subsection{Protection of Edge Majorana Flat Band and Bulk Nodes} \label{sec:appendix-majorana-flat-band}
We briefly detail the protection, at the level of non-interacting band theory, of the Majorana cones found in the nodal $p_y$ stripe phase and the Majorana flat bands in both the nodal and gapped $p_y$ stripe phases of the period-two case via reflection and particle-hole symmetries \cite{Chiu2014}. From Fig. \ref{fig:period-2-mf-configs}b, we see that the unit cell of this striped phase consists of two sites, which we label as $a$ (white, low density) sites and $b$ (black, high density) sites. Using this notation, we can write $H_F$ in the usual BdG form (dropping constant terms),
\begin{align}
H_F &= \frac{1}{2} \sum_{\bm k} \Psi_{\bm k}^\dg h(\bm k) \Psi_{\bm k} = \frac{1}{2} \sum_{\bm k} \Psi_{\bm k}^\dg \begin{pmatrix}
h_0(\bm k) & \Delta(\bm k) \\
\Delta(\bm k)^\dg & -h_0(-\bm k)^*
\end{pmatrix} \Psi_{\bm k} \label{eqn:bdg-hamiltonian}
\end{align}
where we have defined the Nambu spinor
\begin{align}
\Psi_{\bm k} = \begin{pmatrix}
a_{\bm k} &
b_{\bm k} &
a^\dg_{-\bm k} &
b^\dg_{-\bm k}
\end{pmatrix}^T
\end{align}
and 
\begin{align}
h_0(\bm k) &= \begin{pmatrix}
2t \cos(k_y) + 4g\rho_b - \mu && -2t \cos(k_x) \\
-2t\cos(k_x) && -2t \cos(k_y) + 4g\rho_a - \mu
\end{pmatrix},\\
\Delta(\bm k) &= \begin{pmatrix}
2i\Delta_a \sin(k_y) && 0 \\
0 && 2i\Delta_b \sin(k_y)
\end{pmatrix}.
\end{align}
Here, $\rho_{a,b}$ are the average densities on the $a$ and $b$ sites and $\Delta_{a,b} > 0$ the pair fields on the links connecting the $a$ and $b$ sites, respectively. As a BdG Hamiltonian, Eq. (\ref{eqn:bdg-hamiltonian}) automatically satisfies a particle-hole symmetry:
\begin{align}
C h(\bm k) C^{-1} = - h(-\bm k)
\end{align}
with
\begin{align}
C = K \sigma^0 \tau^x \implies C^2 = 1.
\end{align}
Here, $K$ is the complex conjugation operator, the $\sigma^a$, $a=0,\dots,3$, are the Pauli matrices acting on the band index (with $\sigma^0 = 1$), and the $\tau^a$ are Pauli matrices acting on the particle-hole sector. 

The Hamiltonian is also invariant under reflection about the $y$-axis, under which
\begin{align}
\alpha_{x,y} \to \alpha_{-x,y} \implies \alpha_{k_x,k_y} \to \alpha_{-k_x , k_y} \quad (\alpha =a,b).
\end{align}
Eq. (\ref{eqn:bdg-hamiltonian}) thus satisfies the reflection symmetry
\begin{align}
R^{-1} h(k_x, k_y) R = h(-k_x, k_y)
\end{align}
where, since we are dealing with spinless fermions, $R=1$. Defining the composite operator \cite{Chiu2014} $\tilde{C} = RC$, we have that
\begin{align}
\tilde{C}^{-1} h(k_x, k_y) \tilde{C} = -h(k_x, -k_y).
\end{align}
Hence, for fixed $k_x$, $h(k_x, k_y)$ describes a \textit{one-dimensional} (particle-hole symmetric) BdG Hamiltonian in symmetry class $D$, for which we can define the usual $\mathbb{Z}_2$ invariant, $\mathcal{M}(k_x) = \pm 1$ \cite{Kitaev2001}. The nodal points of $h(k_x, k_y)$ then correspond to critical points separating regions in $k_x$ space with different $\mathcal{M}(k_x)$. Since $\mathcal{M}(k_x)$ takes discrete values, this means the nodal points cannot be gapped out by (local) perturbations preserving reflection symmetry. Additionally, if one imposes open boundary conditions in the $y$-direction, a $k_x$ point with $\mathcal{M}(k_x)=-1$ will possess a MZM. Hence, the regions in $k_x$ space with $\mathcal{M}(k_x)=-1$ will yield the observed edge Majorana flat bands. We note that, although $H_F$ also possesses a time-reversal symmetry, we restrict ourselves to a consideration of the $C$ and $P$ symmetries, as they are sufficient to protect the single pair of nodes and non-degenerate Majorana flat bands in the period-two stripe phases.

We can compute $\mathcal{M}(k_x)$ using the usual Pfaffian expression \cite{Kitaev2001}, which requires us to express Eq. (\ref{eqn:bdg-hamiltonian}) in a Majorana basis. Following \cite{Budich2013}, we can define Majorana operators as 
\begin{align}
\begin{pmatrix}
\gamma^a_{\bm k, 1} \\
\gamma^b_{\bm k, 1} \\
\gamma^a_{\bm k, 2} \\
\gamma^b_{\bm k, 2} 
\end{pmatrix} = \sqrt{2} U \Psi_{\bm k}, \qquad 
U = \frac{1}{\sqrt{2}} \begin{pmatrix}
\sigma^0 && \sigma^0 \\
-i\sigma^0 && i\sigma^0
\end{pmatrix}
\end{align}
where $\sigma^0 = 1$ acts on the band index. In the Majorana basis the BdG Hamiltonian takes the form
\begin{align}
iA(\bm k) = U h(\bm k) U^\dg,
\end{align}
where
\begin{align}
A(\bm k) &=\begin{pmatrix}
 0 &  q(\bm k)    \\
 -q^\dg(\bm k) & 0 
\end{pmatrix} , \quad 
q(\bm k) = h_0(\bm k) + \Delta(\bm k).
\end{align}
Note that $A(\bm k)$ is antisymmetric for $k_y = 0 , \pi$. We have that \cite{Kitaev2001}
\begin{align}
\mathcal{M}(k_x) = \sgn\left[ \text{Pf}( A(k_x,k_y=0)) \text{Pf}( A(k_x,k_y=\pi)) \right]
\end{align}
where Pf$(M)$ is the Pfaffian of the antisymmetric matrix $M$. Explicitly,
\begin{align}
\begin{split}
\mathcal{M}(k_x) = &\sgn\left[(4t^2 \cos^2 k_x + (2t-4g\rho_b + \mu)(2t+4g\rho_a -\mu)) \right. \\
					&\times \left. (4t^2 \cos^2 k_x + (2t-4g\rho_a+\mu)(2t+4g\rho_b-\mu)) \right].
\end{split} \label{eqn:z2-invariant}
\end{align}
In Fig. \ref{fig:majorana-flatband-dispersions}c, we have plotted $\mathcal{M}(k_x)$ on top of the BdG spectra with periodic boundary conditions in both directions as a horizontal line at $E=0$. when $\mathcal{M}(k_x) = -1 \, (+1)$, the line is purple (yellow). We see that the changes in $\mathcal{M}(k_x)$ coincide exactly with the projected positions of the bulk nodes and the Majorana flat bands, shown in Fig. \ref{fig:majorana-flatband-dispersions}b, exist in regions with $\mathcal{M}(k_x) = - 1$, up to energy splittings due to the finite size of the system.

\subsection{Majorana Zero Modes at Lattice Dislocations} \label{sec:appendix-dislocations}
We claimed in the main text that lattice dislocations of the gapped period-two stripe phase should trap MZMs. To see this, we use the results of the previous subsection and the argument of Ref. \cite{Mishmash2019}. Let us consider the gapped stripe phase on a torus with $L_{x,y}$ unit cells in the $x$ and $y$ directions, respectively. We can view the torus as a one dimensional system of length $L_y$, consisting of a set of $L_x$ coupled wires. Here, each wire is a two-leg ladder with the two legs consisting of the $a$ and $b$ sites defined in the previous subsection (see Fig. \ref{fig:period-2-mf-configs}b). Each wire can be characterized by the usual $\mathbb{Z}_2$ invariant, $\mathcal{M}=\pm 1$. The torus, viewed as a one-dimensional system, is thus characterized by a $\mathbb{Z}_2$ invariant of $\mathcal{M}^{L_x}$. In order to determine the value of $\mathcal{M}$, we simply need to compute the $\mathbb{Z}_2$ invariant of the torus system with $L_x=1$. This amounts to computing $\mathcal{M}(k_x=0)$ [see Eq. \eqref{eqn:z2-invariant}], which, as shown in Fig. \ref{fig:majorana-flatband-dispersions}c, is indeed $-1$ in the gapped stripe phase. We can thus interpret the period-two gapped stripe phase as an array of topologically non-trivial Kitaev chains, implying that lattice dislocations (along the direction of the stripes) will trap MZMs, as claimed.

\bibliography{FCIreferences}

\end{document}